\documentclass[article,aps,prd,12pt,notitlepage]{revtex4-1}
\usepackage{amsmath,graphicx}
\usepackage{bm}
\usepackage{amssymb}
\usepackage{graphicx,xcolor}
\usepackage{epigraph}
\usepackage{csquotes}
\usepackage{amsmath,amssymb,stmaryrd}
\usepackage{dsfont}
\usepackage{braket}
\begin{document}
\def\e{\enquote}
\def\la{{\langle}}
\def\ra{{\rangle}}
\def\a{{\alpha}}
\def\q{\quad}
\def\ta{t_0}
\def\w{\tilde}
\def\om{\omega}
\def\t{\tilde{t}}
\def\a{\hat{A}}
\def\H{\mathcal{H}}
\def\N{\mathcal{N}}
\def\h{\hat{H}}
\def\E{\mathcal{E}}\def\la{{\langle}}
\def\u{\hat U}
\def\U{\hat U}
\def\C{\hat C}
\def\D{Q}
\def\S{\tilde S}
\def\A{{\mathcal A}}
\def\AA{{\tilde A}}
\def\AAA{{\hat A}}
\def\B{{\hat  B}}
\def\Delt{\tilde \Delta}
\def\QQ{\hat S}
\def\ppi{\hat \pi}
\def\R{\text {Re}}
\def\I1{\text {Im}}
\def\e{\enquote}
\def\qq{s}
\def\up{\uparrow}
\def\do{\downarrow}
\def\Q{\hat Q}
\def\fb{\overline F}
\def\wb{\overline W}
\def\nl{\newline}
\def\h{\hat H}
\def\Pd{Pad\`e }
\def\Pdm{\text{Pad\accute{e} }}
\def\ff{\overline q}
\def\k{\overline k}
\def\F {Q}
\def\f{q}
\def\lm{\lambda}
\def\lmu{\underline\lambda}
\def\q{\quad}
\def\t{\tau}
\def\f{\overline f}
\def\l{\ell}
\def\n{\\ \nonumber}
\def\ra{{\rangle}}
\def\jv{{\vec R}}
\def\Rv{{\bf R}_{\text {A}\gets \text{BC}}}
\def\Rf{{\bf R}_{\text {F}\gets \text{HH}}}
\def\Ep{{\mathcal{E}}}
\def\T{T_{total}}
\def\M{{\mathcal{M}}}
\def\omga{{\epsilon}}
\def\+{\text{up}}
\def\-{\text{down}}
\def\h{\hat{H}}
\def\b{\color{black}}
\def\r{\color{black}}
\def\r{\color{red}}
\def\m{\Omega}
\def\f{\text{f}}
\def\ii{\text{i}}
\def\nuu{\nu_{\text {i}}}
\def\nui'{\nu_{\text {f}}}
\def\I{\text{I}}
\def\II{\text{II}}
\def\III{\text{III}}
\def\IV{\text{IV}}
\newcommand{\Thet }{\Theta }
\newcommand{\pd }{Pad\'{e} }
\newcommand{\PAD }{Pad\'{e}\quad}
\newcommand{\PD }{$\text{Pad\acute{e}}$}
\newcommand{\lla }{\leftarrow }
\newcommand{\va }{\varphi }
\newcommand{\K }{m}
\newcommand{\Ip }{\text{Im}}
\newcommand{\tr }{\theta_{\text{R }}}
\newcommand{\Om }{\Omega }
\newcommand{\nn }{{\nu_{\text{f}}\gets \nu_{\text{i}}}}
\newcommand{\x }{\textbf }
\renewcommand{\thefootnote}{\fnsymbol{footnote}}
\title{
%Complex angular momentum theory of state-to- state integral differential cross sections: resonance effects in the 
%A single resonance, responsible for the unusual behaviour of angular distributions of the 
A transition state resonance radically reshapes angular distributions of the
$F+H_2 \to  FH( v_f=3)+H$ reaction in 
the  $62-102$ $\text{meV}$ energy range }
\author{Dmitri Sokolovski$^{a,b,c}$}
\email{dgsokol15@gmail.com}
\author {Dario De Fazio $^{d}$}
\author {Elena Akhmatskaya $^{b,e}$}
%\author {Chengkui Xiahou $^{f}$}
%\author { J. N. L. Connor $^{g}$}
%\author{D. Alonso$^{d}$}
%\author{S. Brouard$^{d}$}
\affiliation{\footnotesize $^{a}$ Departmento de Qu\'imica-F\'isica Qu\'imica-F\'isica, Universidad del Pa\' is Vasco, UPV/EHU, 48940, Leioa, Spain}
\affiliation{$^{b}$ IKERBASQUE, Basque Foundation for Science, Plaza Euskadi 5, 48009, Bilbao, Spain}
\affiliation{$^{c}$ EHU Quantum Center, Universidad del Pa\'is Vasco, UPV/EHU, 48940, Leioa, Spain}
\affiliation{$^d$ Istituto di Struttura della Materia-Consiglio Nazionale delle Ricerche, 00016 Roma, Italy}
\affiliation{$^e$ Basque Center for Applied Mathematics (BCAM), Alameda de Mazarredo 14, 48009, Bilbao, Spain}
%\affiliation{$^f$ School of Pharmacy, Qilu Medical University, Zibo Economic Zone, Zibo City 255300, Shandong, People's Republic of China}
%\affiliation{$^g$  Department of Chemistry, The University of Manchester, Manchester M13 9PL, United Kingdom}
 \date\today
%\begin{document}
%\author[1,2,3,*]{D. Sokolovski}
%\affil[1]{Departmento de Qu\'imica-F\'isica, Universidad del Pa\' is Vasco, UPV/EHU, 48940, Leioa, Spain}
%\affil[2]{IKERBASQUE, Basque Foundation for Science, Plaza Euskadi 5, 48009, Bilbao, Spain}
%\affil[3]{EHU Quantum Center, Universidad del Pa\' is Vasco, UPV/EHU, 48940, Leioa, Spain}
%\affil[*]{corresponding author.dgsokol15@gmail.com}
%\begin{document}
%\affiliation{$^c$ 
%Department of Particle Physics and Astrophysics, Weizmann Institute of Science, Rehovot, 76100, Israel}
\begin{abstract}
\begin{center}
{\bf ABSTRACT}
\end{center}
\noindent
%{ { \textcolor{red}
{Reactive angular distributions of the benchmark $F+H_2(v_{\text i}=0) \to FH(v_{\text f}=3)+H$ reaction show unusual 
propensity towards small scattering angles, a subject of a long debate in the literature. We use Regge trajectories to quantify the resonance contributions
to state-to-state differential cross sections. Conversion to complex energy poles allows us to attribute the effect
almost exclusively to a transition state resonance, long known to exist in the $F+H_2$ system and its isotopic variant 
$F+HD$.  For our detailed analysis of angular scattering we employ the package \texttt{DCS\_Regge}, recently developed for the purpose [{\it Comp. Phys. Comm.}, 2022, {\bf 277}, 108370.]}
\end{abstract}
%\pacs{03.65.Ta, 03.65.AA, 03.65.UD}
%{Foundations of quantum mechanics}
%\pacs{03.65.AA}{Quantum systems with finite Hilbert space}
%\pacs{03.65.UD}{Entanglement and quantum nonlocality}
\maketitle
%\end{document}
%\section{Introduction}
Reactive differential cross sections (DCS) are known to be sensitive to the details of reaction mechanism and are, therefore, a source of useful information.
A simple picture of a collision between an atom an a diatomic molecule is as follows (see, e.g., \cite{OPT}).
In atom-diatom reaction, transfer of an atom is more likely when the distance between the collision partners is short, i.e., 
for small angular momenta, or small impact parameters. With the forces acting between the atoms being of predominantly repulsive nature, 
a rapid encounter between the reactants is likely to favour large reactive scattering angles $\tr$. Such a direct mechanism can, therefore, 
be expected to produce a backward-peaked reactive DCS,  rapidly decreasing as $\tr$ tends to $0$.
\newline
This is, however, not what one observes  for certain transitions of the benchmark reaction $F+H_2\to FH+H$ \cite{OBS0}, \cite{OBS}, \cite{CAN}. 
%(Although the common notation of fluorine is $HF$, we will continue 
Contrary to the above expectation, the DCS for rovibrational manifolds $v_\ii=0, j_\ii=0 \to v_\f=3,j_\f=0,1,2$ exhibit a high forward 
($\tr=0$) peak, followed by pronounced oscillations which affect the entire angular range $0^o\le \tr \le 180^o$. 
%\newline
The question whether resonances can  lead to observable effects in the $F+H_2$ reaction has long  been debated in the literature \cite{CAN}, 
with opposing views expressed by the authors of Refs.
% Refs.
   \cite{NORES1}-\cite{NORES6} and 
  \cite{OBS}, \cite{RES1}-\cite{RES5}. 
Further readings on the subject can be found in a recent review \cite{REV}  and Refs. therein. 
 \newline
  Reactive DCS can be expected to bear witness of resonance behaviour \cite{OBS} and deserve, therefore, 
  a careful examination. 
%%  affirming their importance. 
%The question is best answered 
The question is best settled 
by an analysis, capable of both quantifying an effect and unambiguously linking it to a resonance known to exist for the $F+H_2$ system. 
Recently, a study of three $F+H_2\to FH+H$ zero-helicity transitions, $\m_\ii=\m_\f=0$, at a translational energy of $62.09$ meV, related the unusual behaviour of the state-to-state reactive DSC to the presence of a single Regge pole \cite{PAPERI}. 
%\newline
The purpose of this paper is to confirm this hypothesis and complete the analysis of Ref.\cite{PAPERI} by extending it to a broader energy range $62.09-101.67$ $\text{meV}$.
We obtain the relevant Regge trajectories and, by converting Regge poles into poles in the complex energy plane, 
attribute the effect to just one transition state resonance \cite{RES2}, known to exist for the $F+H_2$ system and its isotopic variant $F+HD$ (see, for example, 
 Refs.\cite{A2}-\cite{FHD2}). 
%In this way we are able to quantify the contributions 
\section{Methods}
For our analysis we will rely on a methodology somewhat different from those used in \cite{PAPERI}.
It was recently implemented in the software \texttt{DCS\_Regge}, 
now available in the public domain \cite{CPC3}. The method is discussed in detail in  \cite{CPC3}, and here we only repeat what is necessary for
the present narrative. A differential cross section (DCS), $\sigma_{\nn}$, at an angle $\tr$ and an energy $E$ is given by an absolute square of the scattering amplitude,
 \begin{eqnarray}\label{1}
 \sigma_{\nn}(\tr,E) = | f_{\nn}(\tr,E)|^2,
 \end{eqnarray}
 where 
the composite indexes $\nu_\ii=(v_\ii,j_\ii,\m_\ii)$ and $\nu_\f=(v_\f,j_\f,\m_\f)$
include the initial and final vibrational ($v$), rotational ($j$) and helicity ($\m$) quantum numbers of the system. 
In the zero-helicity case,  $\m_\ii=\m_\f=0$, the amplitude is given by a simple partial wave sum,
 \begin{eqnarray}\label{3}
 f_{\nn}(\tr,E) = ( ik_\nu)^{-1}\sum_{J=0}^\infty
 (J+1/2)
 S^J_\nn(E) P_J(\cos(\pi -\tr)),
 \end{eqnarray}
where $J$ is the total angular momentum quantum number,  $S^J_\nn$ is a body-fixed scattering matrix element, $k_\nu \equiv k_{v_i,j_\ii}$ is the initial translational wave vector of the reactants, and  $P_J(\bullet)$ is Legendre polynomial (see, e.g., \cite{BRINK}). In practice, the sum is terminated at some $J=J_{max}>>1$, and this inequality 
holds under the semiclassical condition, assumed throughout the rest of the paper.  
\newline
{For a chosen state-to-state transition, the code $\texttt{DCS\_Regge}$ evaluates two \e{unfolded} amplitudes (the  usual shorthand notation 
$\lambda \equiv J+1/2$ is used below), 
 \begin{eqnarray}\label{4}
\tilde f (\va,E)=\int_{0}^\infty \sqrt \lm  S_\nn(\lambda,E)\exp(i\lm \va)d\lm,\n
\tilde g (\va,E)=\int_{0}^\infty \lm  S_\nn(\lambda,E)\exp(i\lm \va)d\lm, 
 \end{eqnarray}
both are functions of the newly introduced  angular variable $\varphi$, which varies between $-\infty$ and $\infty$.
The function $S_\nn(\lambda,E)$ in (\ref{4}), such that $S_\nn(J+1/2,E)=S^J_\nn(E)$, $J=0,1,...J_{max}$, is the analytic continuation
of the $S$-matrix element into the complex angular momentum (CAM) plane, achieved by means of
Pad\'e  { approximation} (more details are given in \cite{CPC1}). 
The scattering amplitude (\ref{3}) for $\pi/J_{max}\lesssim \tr \lesssim \pi-\pi/J_{max}$  can now be obtained by \e{folding back} the amplitudes (\ref{4}) \cite{CPC3}, 
\begin{eqnarray}\label{5}
f_\nn(\tr,E)= %\q\q\q\q\q\q\q\q\q\q\q\n
%(ik)^{-1}[2\pi \sin \tr]^{-1/2}\sum_{m=-\infty}^\infty\tilde f(\varphi_m)\exp(-i\pi/4 -im\pi/2),\n
(ik_\nu)^{-1}[2\pi \sin \tr]^{-1/2}\sum_{m=-\infty}^\infty\tilde f(\varphi_m)\exp(-i\pi/4 -im\pi/2)\q\q\q\n
\equiv \sum_{m=-\infty}^\infty f^{(m)}_\nn(\tr,E),\q\q
 \end{eqnarray}
where 
%\q\q\q\q\q\n
\begin{eqnarray}\label{5a}
\varphi_m(\tr) \equiv (-1)^{m+1}\tr + \pi[m+1/2+(-1)^m/2].
\q\q\q\q\q\q\q\q
 \end{eqnarray}
For $\tr=0$, and $\tr=\pi$ one has
\begin{eqnarray}\label{6}
 f_{\nn}(\tr=0,E)=
 %\q\q\q \q\q\q\n 
 -k_\nu^{-1}\sum_{\K=-\infty}^\infty(-1)^\K 
\tilde g((2\K+1)\pi)\equiv \sum_{m=-\infty}^\infty f^{(m)}_\nn(0,E),\q\n
%\end{eqnarray}
%begin{eqnarray}\label{8a}
  f_{\nn}(\tr=\pi,E)=
  %=\sum_{\K=-\infty}^\infty f^{BW}_{\nu^{\prime} \gets \nu}(E|M)=}
  %  \q\q\q \q\q\q\n
   (ik_\nu)^{-1}
  \sum_{\K=-\infty}^\infty(-1)^\K\tilde g(2\K\pi)\equiv \sum_{m=-\infty}^\infty f^{(m)}_\nn(\pi,E).\q\q
 % f^{FS}_{\nu^{\prime} \gets \nu}(\tr,E)=( ik_\nu)^{-1}[2\pi \sin(\tr)]^{-1/2}\sum_{\K=0}^\infty \tilde f (\va^{FS}_\K)
 \end{eqnarray}
The rationale behind Eqs.(\ref{5})  becomes clear by considering the classical limit of atom-diatom reaction 
$\text{A+BC}\to \text{AB+C}$. The variable of interest is the winding angle $\varphi$, i.e., the angle swept in the course of collision
by the projection of the Jacobi vector $\Rv$, drawn from the centre of mass of the $\text{BC}$ pair to the atom $\text{A}$, onto the plane perpendicular to the total angular momentum $\bf J$ (see Supplementary Material $A$). 
One can attempt to achieve a semiclassical description by ascribing a probability amplitude to the reactive trajectories leading 
to all winding angles $\varphi_m(\tr)$, consistent with the scattering angle $\tr$ [cf. Eq.(\ref{5a})], and adding up the amplitudes
 in accordance with the basic rule of quantum mechanics. The correct formula (\ref{5})  contains, however, 
additional phase factors $\exp(-i\pi/4 -im\pi/2)$, and fails for both small and large scattering angles.
 The reason for this is the coalescence
of winding angles $\pi -\tr +2m\pi$ and $\pi+\tr+2m\pi$  as $\tr \to 0$, and of $\pi -\tr +2m\pi$, and $\pi+\tr+2(m-1)\pi$  as $\tr \to \pi$.
With the two angles no longer distinguishable,  Eq.(\ref{5}) must be replaced by one of the Eqs.(\ref{6}). (A detailed discussion of Eqs.(\ref{6})
can be  found in \cite{PCCP}). 
\newline
The analysis proceeds by examining the shapes of the two $\tilde f (\va,E)$ and $\tilde g (\va,E)$.
For a direct reaction one can expect  an $\tilde f (\va,E)$ essentially limited to the interval $0\le \va \le \pi$, negligible for $\va \approx 0$, 
considerable for $\va \approx \pi$ and, perhaps, having a small extension into the $\va <0$ region due to purely quantum effects. 
For a reaction, passing through formation of one or several intermediate rotating complexes, $\tilde f (\va,E)$ is likely to extend also into the $\va \ge \pi$ zone. 
%which would produce a characteristic interference pattern in the DCS $ \sigma_{\nn}(\tr,E)$. 
%\newline 
This extension is expected to take the form of one or several \e{exponential tails},  which resonance CAM (Regge)  poles at $J_n$, $n=1,2...$,  in the first quadrant of the CAM plane contribute to the integrals 
in Eqs.(\ref{4}) \cite{PCCP},  
\begin{eqnarray}\label{7}
\tilde f (\va\ge \pi) \approx 
%2\pi i \sum_{n=1}^{N_{res}}\sqrt{\lm_n}Res[S_{\nu' \gets \nu}(E,\lambda_n)]\exp(i\lm_n\va),\q\n
2\pi i \sum_{n=1}^{N_{res}}\sqrt{\lm_n}r_n(E)\exp(i\lm_n\va)\equiv \sum_{n=1}^{N_{res}}\tilde f^{tail}_n (\va),\q\n
\tilde g (\va\ge \pi) \approx 
%2\pi i \sum_{n=1}^{N_{res}} {\lm_n}Res[S_{\nu' \gets \nu}(E,\lambda_n)]\exp(i\lm_n\va),\q
2\pi i \sum_{n=1}^{N_{res}} {\lm_n}r_n(E)\exp(i\lm_n\va)\equiv \sum_{n=1}^{N_{res}}\tilde g^{tail}_n (\va).\q
 \end{eqnarray}
In  Eq.(\ref{7}),  $r_n(E) \equiv \lim_{J\to J_n}(J-J_n)S_{\nn}(E,J)$ is the residue of the $S$-matrix element at $J=J_n$,  
 and the sum  is over $N_{res}$  \e{physically significant poles}. The notion is, in this context, self-explanatory.
Significant one is a pole whose contribution to $\tilde f (\va\ge 0)$ and $\tilde g (\va\ge 0)$ is sufficient to cause observable interference effects
in the DCS [cf Eqs.(\ref{5})]. Such a pole must lie not too far from the real $J$-axis (or its tail will be too short), and have  a sufficiently large residue
(or the tail will be too small). 
Finally, we note that Eqs.(\ref{7}) are consistent with a picture of a rotating intermediate triatomic complex which continues to decay into products.
% and at the same rotates.
Triatomic's moment of inertia $I$ and $\R[J_n]$ determine its angular velocity, $\om \approx \R[J_n]/I$, while $1/\text{Im}[J_n]$ yields the typical rotation angle \cite{CPC3}.
\newline
Next we check whether for the chosen transitions the unfolded amplitudes in Eqs.(\ref{4}) have tails described by Eqs.(\ref{6}), and ascribe every found tail to a Regge pole, or poles, of the $S$-matrix element.  Importantly, we will  relate each CAM pole to its complex energy counterpart, and explain the shape of the DCS in terms of the well known $F+H_2$ resonances \cite{A0},\cite{A1}, something not yet done in \cite{PAPERI}. 
%(The reader who wants to know the result now can consult the article's title).
 %%%%%%%%%%%%%%%%%%%
\section{Results} 
As in \cite{PAPERI}, we use the $S$-matrix elements of the title reaction, computed by the hyperspherical method of \cite{CALC}
on the Fu-Xu-Zhang (FXZ) potential energy surface (PES) \cite{FXZ}.

{\it A. The $(0,0,0)\to (3,0,0)$ transition.} The DCS (\ref{1}) shown in Fig.\ref{F3}a, exhibits a high forward scattering peak, whose height varies little 
across the translational energy range $62.09-101.67$ meV considered here. There are also regular oscillations observed, at all energies, in the entire angular range $0\le \tr \le 180^o$. 
A closer inspection reveals  much smaller patterns superimposed on the DCS in the regions $65-71$ meV and  $85-92$ meV (indicated  by arrows in Fig.\ref{F3}a).
To explain this behaviour we employ the \texttt{DCS\_Regge} code of Ref.\cite{CPC3}. 
\begin{figure}[!ht]
\includegraphics[angle=0,width=13cm, height= 13cm]{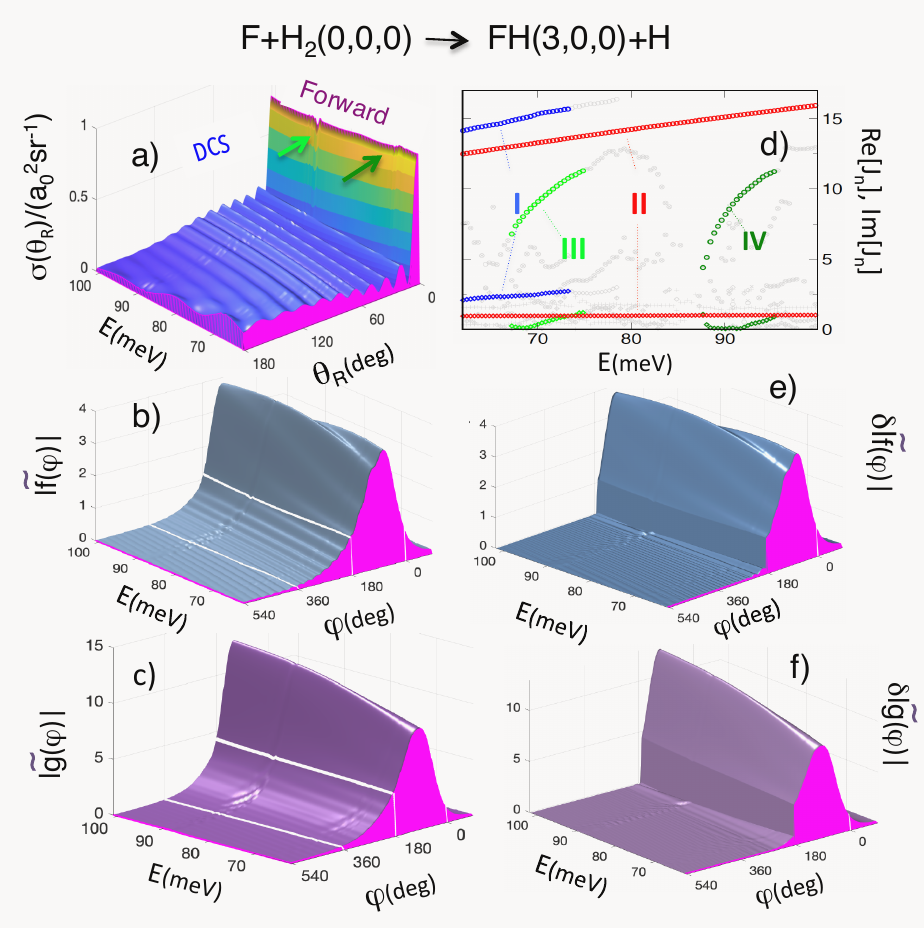}
\caption {a) Reactive differential cross section  $\sigma_{\nn}(\theta_R)$ for  the $v_i=0, j_i=0,\m_\ii=0 $ and  $v_\f=3, j_\f=0,\m_\f=0$ 
vs. $\theta_R$ and $E$. Small patterns which occur in the regions $65-70$ meV and  $85-90$ meV in the entire angular range are indicated by arrows.
\newline
b) The modulus of the unfolded amplitude $\tilde{f}$ vs. the winding angle  $\varphi$ and $E$.
%Patterns which occur in the regions $65-70$ meV and  $85-90$ meV are indicated by arrows. 
\newline
c) Similar to (b), but for the unfolded amplitude $\tilde{g}$. 
\newline
d) Real (circles) and imaginary (diamonds) parts of the four Regge trajectories.
\newline
e) The difference between $|\tilde f(\varphi)|$ and $|\tilde f_{\II}^{tail}(\varphi)|$ [cf. Eqs.(\ref{4}) and (\ref{7}) ]
subtracted for  $\varphi > 180^o$.
\newline
f) The difference between $|\tilde g(\varphi)|$ and $|\tilde g_{\II}^{tail}(\varphi)|$ [cf. Eqs.(\ref{4}) and (\ref{7}) ]
subtracted for  $\varphi > 180^o$.}
\label{F3}
\end{figure}
\newline 
Both unfolded amplitudes in Eq.(\ref{4}), $\tilde f (\va,E)$ and $\tilde g (\va,E)$, shown in Fig.\ref{F3}b and c, respectively, are smooth functions of $\va$ and $E$, and have tails which extend 
into $\va\ge 180^o$ region and become negligible at $\va \sim 500^o$. The oscillations in Fig.\ref{F3}a must, therefore, result from interference between the 
part of $\tilde f $ contained in the region  $0\le \va\le 180^o$, and the tail, which we expect to be produced by capture into a metastable state, or states 
[cf. Eqs.(\ref{5}) and (\ref{7})].
\newline
To check whether this is the case, we plot the Regge poles in Fig.\ref{F3}d. There are four Regge trajectories, labelled by Roman numerals, $n=\text{I, II, III, IV}$.
For $E=62.09$ meV, trajectory $\II$ contains the pole at $J=12.49+0.95i$, previously found in \cite{PAPERI}, and we expect it to be the most significant of the four. Indeed, in Fig.\ref{F3}e the difference 
$\delta|\tilde f (\va,E)|\equiv |\tilde f (\va,E)|-|\tilde f^{tail}_{\II} (\va,E)|\chi(\va-\pi)$, where $\chi(x)=1$ for $x\ge0$ and $0$ otherwise, practically vanishes for $\va \ge 180^o$, 
which shows that  the pole $\II$ responsible for almost all of $\tilde f(\va,E)$ $\va \ge 180^o$ region. 
The same is true for the difference  $\delta|\tilde g(\va,E)|\equiv |\tilde g (\va,E)|-|\tilde g^{tail}_{\II} (\va,E)|\chi(\va-\pi)$, shown in Fig.\ref{F3}f.
Thus, there can be little doubt that the tails of both unfolded amplitudes and, therefore, the oscillations in the DCS in Fig.\ref{F3}a, are largely caused by the resonance II 
%$\tilde g (\va,E)$
 [cf. Fig.\ref{F3}f]. 
\newline
An inspection of the pole positions in Fig.\ref{F3}d and  the corresponding residues in Fig.\ref{F4} explains why other poles can have only minor effect on 
$ \sigma_{3,0,0\gets 0,0,0}(\tr,E)$.
 The resonance  $\I$ (i.e. the Regge pole  $\I$) has the largest residue, but 
also a large imaginary part.  For $\va \ge \pi$, its tail, reduced by a factor $\exp\{ - \text{Im}[J_\I]\va\}$, is both short and small. (It needs, however, to be taken into account 
when calculating the forward scattering cross section at lower energies, as will be shown shortly.) The residues of the poles $\III$ and $\IV$ are very small, 
and even though both resonances are long-lived [cf. Fig.\ref{F3}d], they are accountable only for the small patterns indicated by arrows in Fig.\ref{F3}a.
%%%%%%%%%%%%%%%%%%%%%%%%%%%%%%%%%
\begin{figure}[h]
\includegraphics[angle=0,width=12cm, height= 7cm]{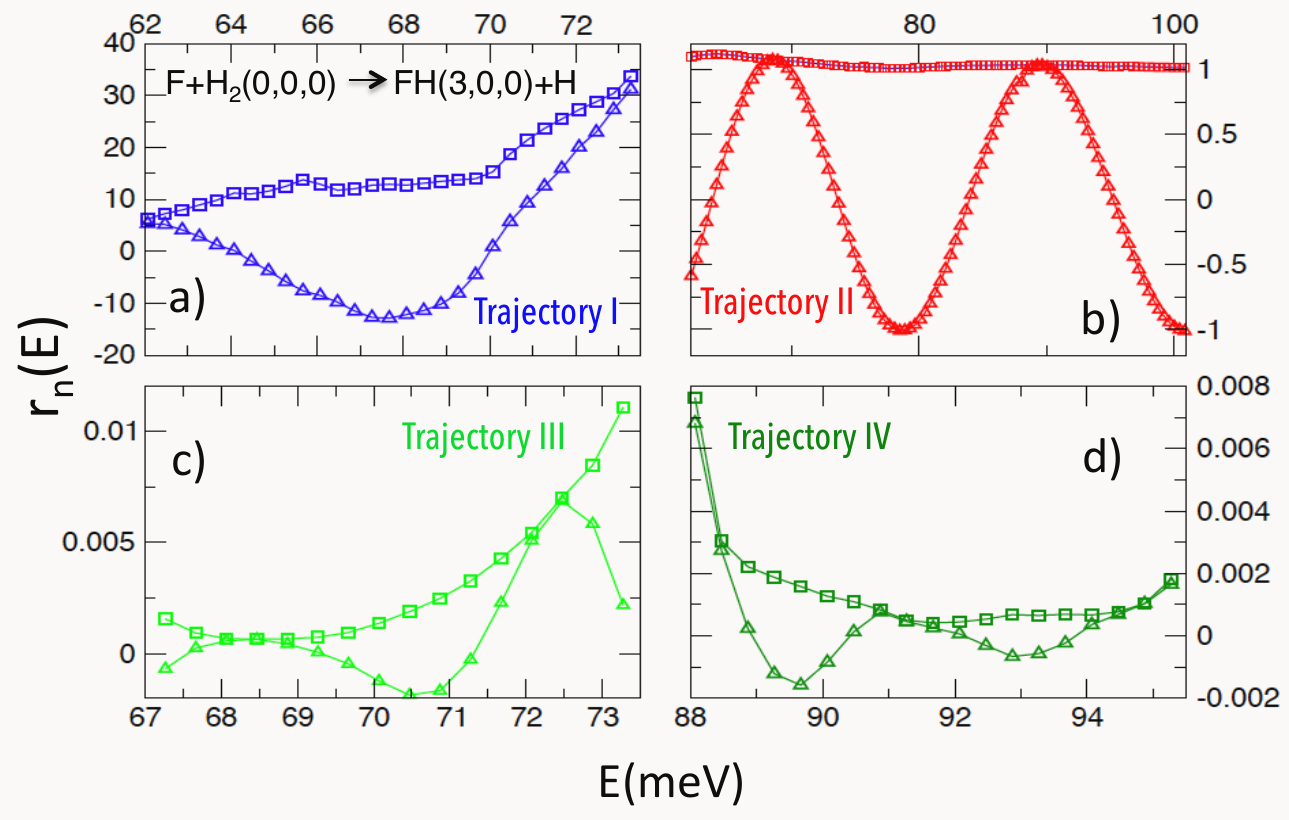}
\caption {Moduli (squares) and real parts (triangles) of the residues $r_n(E)$,  $n=\text{I, II, III, IV}$,  for the four Regge trajectories in Fig.\ref{F3}d. }
\label{F4}
\end{figure}
%%%%%%%%%%%%%%%%%%%%%%%%%%%%
\newline
To quantify the analysis, we consider the energy dependence of the DCS at a chosen scattering angle, 
first by identifying the important terms in the expansions (\ref{5})-(\ref{6}), using, where possible, approximations (\ref{7}), 
and comparing the result with the exact DCS.  
 %%%%%%%%%%%%%%%%%%%%%%%%%%%-----------------
\begin{figure}[!ht]
\includegraphics[angle=0,width=15cm, height= 6cm]{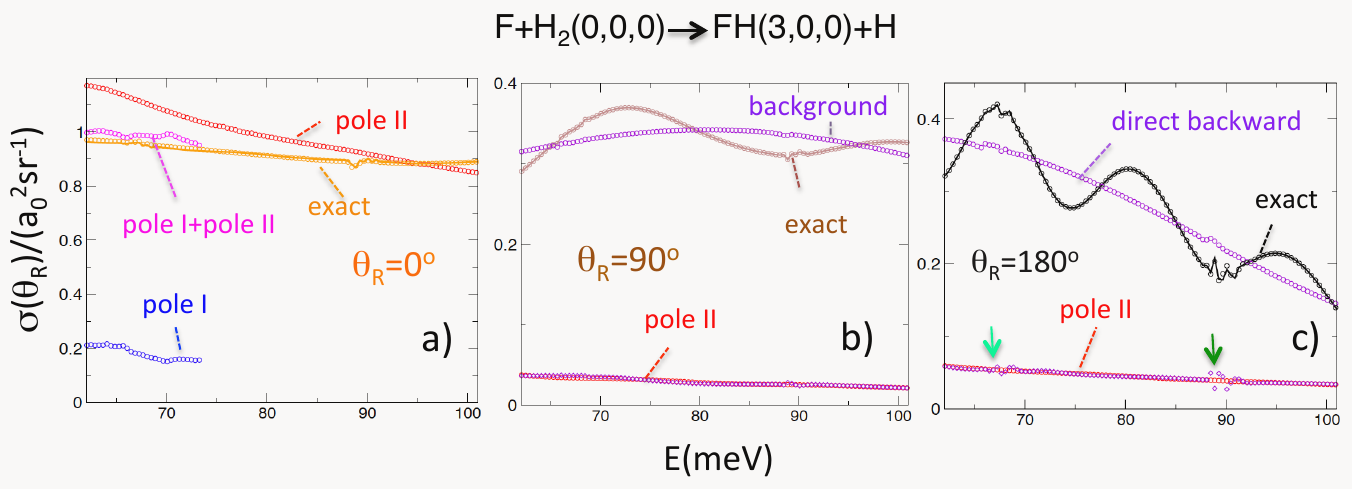}
\caption {a) Exact DCS (\ref{1}) at $\tr=0^o$ (solid orange) and $\left |f^{(0)}_\nn(0,E)\right |^2$ (orange circles). 
Also shown are
 the resonance contributions $\left |k_\nu^{-2} \tilde g^{tail}_{\I} (\pi)\right |^2$ (blue circles), 
$k_\nu^{-2}\left | \tilde g^{tail}_{\II} (\pi)\right  |^2$, as well as a coherent sum of the two,  $k_\nu^{-2}\left |\tilde g^{tail}_{\I}+ \tilde g^{tail}_{\II} (\pi)\right |^2$ (malva circles).
\newline
b) Exact DCS (\ref{1}) at $\tr=90^o$ (solid brown). Also shown are the background term, $\left |f^{(-1)}_\nn(\pi/2,E)+f^{(0)}_\nn(\pi/2,E)\right |^2$ (violet circles), 
$\left |f^{(1)}_\nn(\pi/2,E)\right |^2$ (violet diamonds), and their coherent sum $\left |\sum_{m=-1}^1f^{(m)}_\nn(\pi/2,E)\right |^2$ (brown circles). Contribution of the resonance II, 
$\left |(2\pi)^{-1/2} k_\nu^{-1} \tilde f^{tail}_{\II} (3\pi/2)\exp(-i\pi/4)\right |^2$,  (red circles) is in good agreement with $\left |f^{(1)}_\nn(\pi/2,E)\right |^2$.
\newline
c) Exact DCS (\ref{1}) at $\tr=180^o$ (solid black). Also shown are the direct term $\left |\tilde f^{(0)}_\nn(\pi,E)\right |^2$ (violet circles), $\left |\tilde f^{(1)}_\nn(\pi,E)\right |^2$ (violet diamonds), and their
coherent sum $\left |\sum_{m=0}^1f^{(m)}_\nn(\pi,E)\right |^2$ (black circles). Contribution of the resonance II, $k^{-2}\left |\tilde g^{tail}_{\II}(2\pi)\right |^2$ (red circles), is  in good agreement with $\left |\tilde f^{(1)}_\nn(\pi,E)\right |^2$, with small discrepancies noted where the resonances III and IV need to be taken into account (as indicated by the arrows). 
}
 \label{F5}
\end{figure}
%%%%%%%%%%%%%%%%%%%%%%%%%%%
\newline
A good approximation to the {\it forward} DCS is obtained by taking into account only the zeroth term in (\ref{6}), $f^{(0)}_\nn(0,E)$ (orange circles in Fig.\ref{F5}a).
One can try to attribute it to the decay of the recently formed resonance II into the forward direction $\tr=0$, 
%and neglecting the process in which $\Rf$ completes another full rotation, 
$f_{\nn}(0,E)\approx 
 -k_\nu^{-1} \tilde g^{tail}_{\II} (\pi)$. The result, shown by red circles in Fig.\ref{F5}a,  can be improved, by about a fifth (malva circles in Fig.\ref{5}a), by including also the 
 contribution from the resonance $\I$ 
 %[cf. Eqs.(\ref{7}) and Fig.\ref{F5}a],
 \begin{eqnarray}\label{8}
  f_{\nn}(0,E)\approx  -k_\nu^{-1} [\tilde g^{tail}_{\I} (\pi)+\tilde g^{tail}_{\II} (\pi)].
   \end{eqnarray} 
 [We are able to follow Regge trajectory $\I$ only until it leaves the region where Pad\'e approximation can be trusted at $E\approx 74$ meV \cite{CPC1}.
The remaining small discrepancy is attributed to the inaccuracy of the asymptote (\ref{7}) if the resonance is formed at a large value of $J$ \cite{PCCP}].

The {\it sideway} DCS at $\tr=90^o$, shown in Fig.\ref{F5}b (solid brown),  is, to an excellent accuracy, a result of interference between the \e{background} term (violet circles), $f^{(-1)}_\nn(-\pi/2,E)+f^{(0)}_\nn(\pi/2,E)$ and $f^{(1)}_\nn(-\pi/2,E)$ (violet diamonds). The latter term is seen to be a result of the decay of the resonance $\II$ (red circles) after $\Rf$ rotates by $3\pi/2$ [cf. Eqs.(\ref{6}) and (\ref{7}) ],
 \begin{eqnarray}\label{9}
  f_{\nn}(\pi/2,E)\approx f^{(-1)}_\nn(-\pi/2,E)+f^{(0)}_\nn(\pi/2,E)
  -\frac{1}{\sqrt{2\pi} k_\nu} \tilde f^{tail}_{\II} (3\pi/2)\exp(-i\pi/4).
    \end{eqnarray} 
(Note that  \texttt{DCS\_Regge} does not distinguish between direct scattering and decay of a resonance for a $\va$ lying between $0$ and $\pi$.
A more sophisticated technique is available \cite{PCCP}, but was deemed too cumbersome to be included into the software.)

 Finally, the oscillations of the {\it backward} DCS (black solid) in Fig.\ref{F5}c  are clearly the result of interference between a direct recoil following a head-on collision, 
 and the decay of the resonance $\II$ after $\Rf$ completes one full rotation [cf. Eqs.(\ref{7})], 
 \begin{eqnarray}\label{10}
  f_{\nn}(\pi,E)\approx  (ik_\nu)^{-1} [\tilde g (0)-\tilde g^{tail}_{\II} (2\pi)]. 
  \end{eqnarray}  
%%%%%%%%%%%%%%%%%%%%%%%%%%

{\it B. Assignment of Regge resonances.} 
%There is curious dualism in describing quantum mechanical resonances in terms of singularities of scattering matrix element $S_\nn(\lambda,E)$, analytic in both the 
%angular momentum and energy variables.
There are two complimentary ways of relating  the same resonance phenomenon to a singularity of a scattering matrix element $S_\nn(\lambda,E)$.
 Fixing a real value of $J$ allows one to look for poles in complex energy (CE) plane, while fixing a value 
of $E$ gives rise to CAM Regge poles. The CE poles,  whose relation to the PES is usually well understood (see \cite{OBS}, \cite{A1},\cite{RES5}),
% (e.g., by the $Q$-matrix method of \cite{A2}-\cite{A3}), 
  are not particularly 
useful for a quantitative analysis of the integral and differential cross sections, given by sums over angular momentum at a fixed value of $E$. 
The Regge poles, for their part, are well suited for such an analysis, but can offer little insight into the dynamics on the PES. 
% into,  e.g., the part of the PES which supports the corresponding metastable state. 
Fortunately,  positions of the poles of one kind can usually be obtained if the positions of poles of the other kind are already known \cite{PCCP2}, so the benefits of both approaches can be combined. 
\newline
In the present case, the task is especially easy since the pole positions of the resonances $\I$ and $\II$ are practically linear functions of $E$ in their respective energy ranges, 
%\begin{eqnarray}\label{11}
$J_{\I,\II}\approx \alpha_{\I,\II} +\beta_{\I,\II} E$. \q
%\q\q\q\q\q\q\q\q\q\q\q\q\q\q\q\q\q\q\q\q\q\n
%\alpha=7.016+0.839i, \q \beta=0.0884+0.00183i.
%\q\q\q\q\q\q\q\q\q
% \end{eqnarray}}
 {Inverting the relation yields the positions of the corresponding CE poles, $E_{\I,\II}$,
\begin{eqnarray}\label{12}
E_{\I,\II}(J)\approx a_{\I,\II}+b_{\I,\II} J, \q 
%\q\q\q\q\q\q\q\q\q\q\q\q\q\q\q\q\q\q\q\q\q\n
%a=-79.511-7.852i, \q b=11.304-0.233i.
%\q\q\q\q\q\q\q\q\q
 \end{eqnarray}}
with the complex constants given by $a_{\I,\II}= - \alpha_{\I,\II}\beta^{-1}_{\I,\II}, \q b_{\I,\II}=\beta^{-1}_{\I,\II}$. 
Numerical fits, shown in Fig.\ref{F6}a, are in good agreement with the exact positions of CE, obtained by Pad\'e
reconstruction of the $S$-matrix element in the complex energy plane. For the $F+H_2$ system, the properties of the CE poles 
have been extensively studied, e.g., in \cite{A1}, albeit on an older and less accurate Stark-Werner (SW) potential surface (Fig.9 of \cite{A1}). 
% slightly different potential surface.
 %\cite{PES1}. 
 After accounting for the expected difference between the FXZ and SW PES (for more detail, see  Supplementary Material $B$),  a comparison in Fig.\ref{F6}b 
 %and Fig.9 of \cite{A1}
attributes the Regge trajectories $\I$ and $\II$ in Fig.\ref{F3}a 
to the well known in the literature resonances  $\text{B}$ and $\text{A}$, respectively. The nomenclature was first introduced in \cite{A0}, and subsequently used by other authors. 
Both A and B are Feshbach resonances, correlated with bound states with support in different regions of the adiabatic potential curve
[cf. Fig.3b of \cite{RES2}].  Resonance $\text{B}$ is a vdW  exit channel resonance, while $\text{A}$, trapped in a deeper well, lies closer to the system's transition state
\cite{RES1}. 
Thus, we find the transition state resonance A, whose tails are clearly visible in Figs.\ref{F3}b,c, to be responsible for the unusual behaviour of the state-to-state DCS 
in Fig.\ref{F3}a. 
 %%%%%%%%%%%%%%%%%%%%%%%
\begin{figure}[!ht]
\includegraphics[angle=0,width=14cm, height= 7cm]{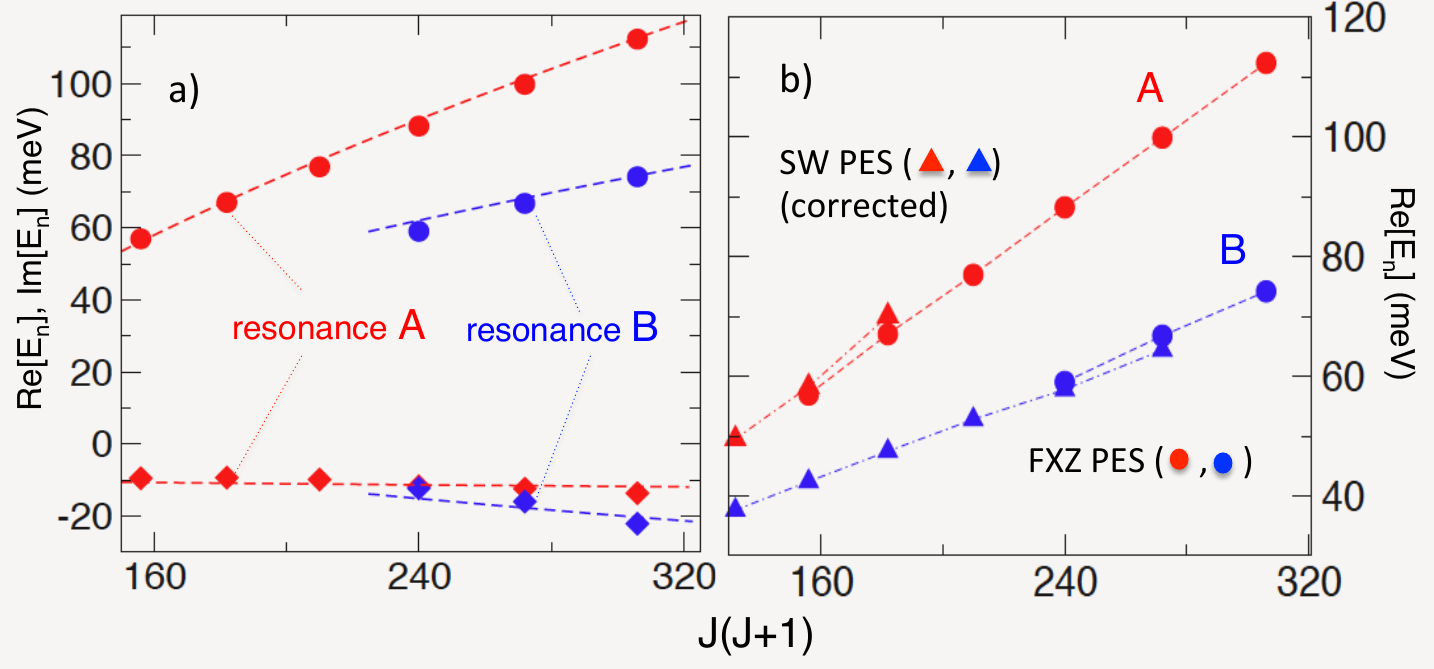}
\caption {
Real (circles) and imaginary (diamonds)  parts of the exact complex 
energy poles, corresponding to the Regge trajectories $\I$ (blue) and $\II$ (red) in Fig.\ref{F3}d.
Also shown by the dashed lines is the approximation (\ref{12}).
The variable $J(J+1)$, rather than $J$, is used to facilitate comparison with Fig.9 of \cite{A1}.
b) Comparison between real parts of the CE poles obtained for the FXZ PES (circles, present work),
and the corresponding poles for the SW PES (used in Ref.\cite{A1}, triangles). The SW results are 
corrected downwards by $13$ meV (see Fig.\ref {F6A} of the Supplementary Material).}
\label{F6} 
\end{figure}
%%%%%%%%%%%%%%%%%%%%%%%%%%%%%%%%%
%%%%%%%%%%%%%%%%%%%%%%%%%%%%%%%%%%%%%

{\it C. Transitions $(0,0,0)\to (3,1,0)$  and  $(0,0,0)\to (3,2,0)$.}
The shapes of the DCSs of these transitions, shown in  Fig.\ref{F7}, are similar to the one in Fig.\ref{F3}a, and can be analysed in 
the same manner (see Supplementary Material $C$). We find that all three transitions share the same Regge trajectories, plotted in Fig.\ref{F8}.
Such a coincidence is to be expected, as the singularities, whether in the CE or CAM plane, are shared by all matrix elements, 
which differ only in the magnitude of the corresponding residues. 
The $(0,0,0)\to (3,2,0)$ transition is, however, different in one respect. 
Its  $\tilde f$- and $\tilde g$-amplitudes  in Figs.\ref{9}c,d reveal an additional  minimum in the $0 \le \va \le \pi$ region
across the whole energy range.
This feature can be traced back to a {\it Regge zero} trajectory shown in Fig.\ref{F10}.  
%%%%%%%%%%%%%%%%%%%%%%%%%%%%%%%
\begin{figure}[!ht]
\includegraphics[angle=0,width=14cm, height= 6cm]{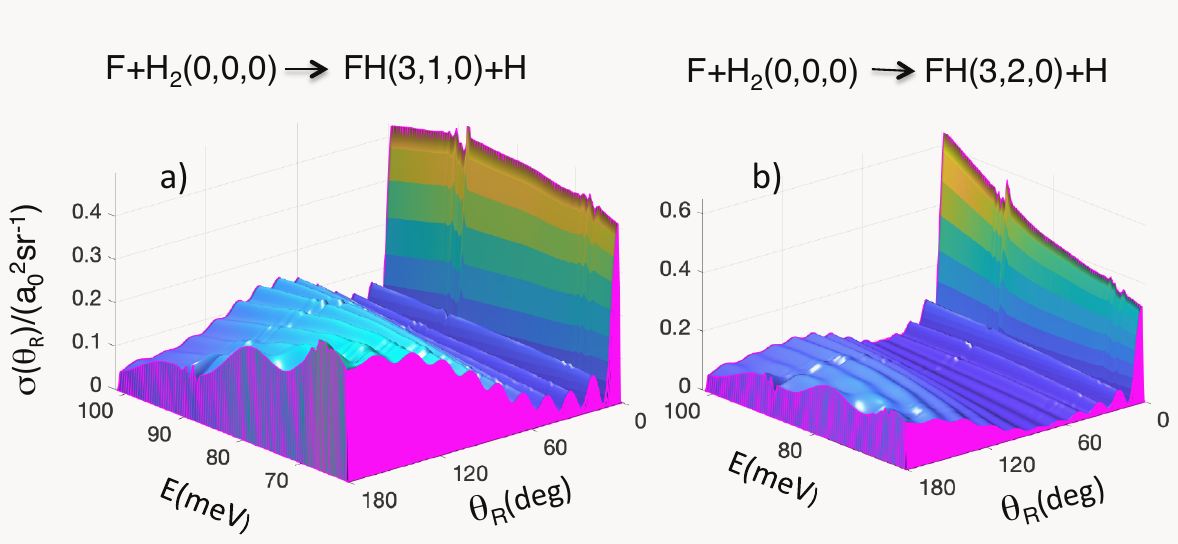}
\caption {a)
Reactive differential cross section  $\sigma_{\nn}(\theta_R)$ for   $\nuu=(0,0,0)$ and  $\nui'=(3,1,0)$ 
%$(3,1,0)$ and $(3,2,0)$ 
vs. $\theta_R$ and $E$.
b) Same as (a), but for $\nui'=(3,2,0)$. } 
\label{F7} 
\end{figure}
%%%%%%%%%%%%%%%%%%%%%%%%%%%%%%%
\begin{figure}[!ht]
\includegraphics[angle=0,width=11cm, height= 7cm]{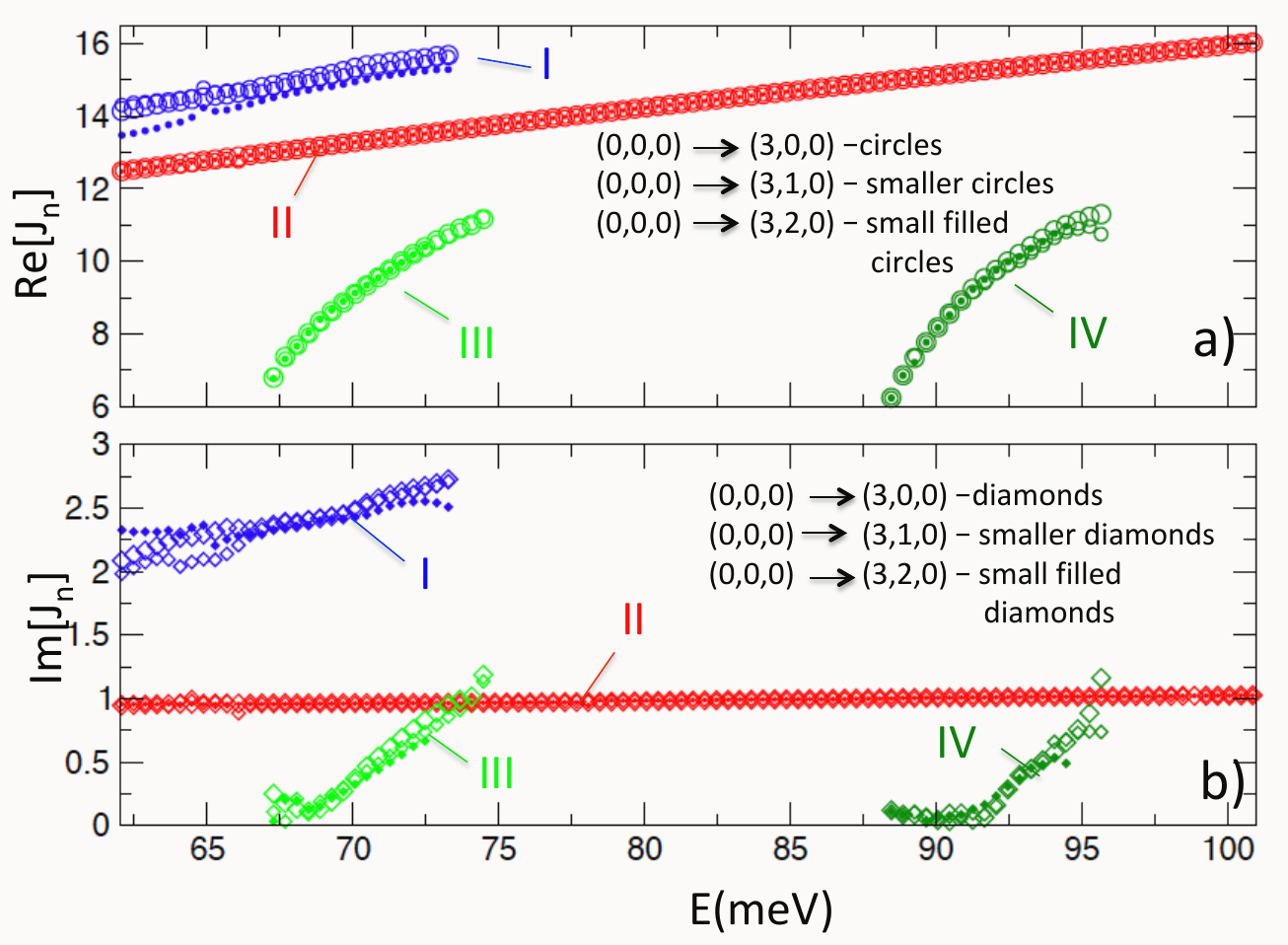}
\caption {a)
Real parts of the Regge trajectories $\I$, $\II$, $\III$, and $\IV$ (circles) for $\nuu=(0,0,0)$ and $\nui'=(3,0,0), 
(3,1,0), (3,2,0)$.
b) Imaginary parts of the same trajectories (diamonds). } 
\label{F8} 
\end{figure}
%%%%%%%%%%%%%%%%%%%%%%%%%%%
%Regge zeroes (see, e.g.,  \cite{DSzero}), i.e., interference minima  of an $S$-matrix element, appear only in multichannel scattering \cite{DSzero}.
%Unlike physical Regge poles, restricted to the first quadrant, the zeroes  may lie anywhere of the complex $J$-plane. 
The zero affecting the $(0,0,0)\to (3,2,0)$ transition is likely the cause of a somewhat poorer agreement with the rainbow theory in Fig.8c of Ref.\cite{PAPERI}.
A more detailed analysis of the $(0,0,0)\to (3,1,0)$  and  $(0,0,0)\to (3,2,0)$ transitions will be given elsewhere \cite{PAPERIII}.
%A zero in the first quadrant 
%(cf. Fig.\ref{FIG9})) has an effect
%opposite to a that of Regge pole, producing minimum  in the 
%reaction probability (plotted vs. $J$), and in both unfolded amplitudes. 

%A zero in the fourth quadrant would produce a dip in the QDF, a  dip 
%in the reaction probability, a well as bumps in both unfolded amplitudes $\tilde f(\varphi)$ and $\tilde g(\varphi)$.
%The presence of the zero in Fig.\ref{FIG9} is responsible, in particular, for a somewhat poorer agreement with the rainbow theory in Fig.8c of Ref.\cite{PAPERI} }
%%%%%%%%%%%%%%%%%%%%%%%%%%%%%%%%%%%%%%
In summary, the same resonance is responsible for the shape of the DCS in Fig.\ref{F3}a as well as in Figs.\ref{F7}a,b. 
\section{Conclusions and discussion}
A thorough analysis revealed that a single resonance is largely responsible for the unusual behaviour of the 
% benchmark 
$F+H_2(0,0,0) \to FH(3,j,0)$ reaction in the entire collision energy range $62.09-101.67$ meV.
This behaviour, we recall, involves a pronounced forward scattering maximum, followed by oscillations
clearly visible at all scattering angles [cf. Figs.\ref{F3} and \ref{F7}].
The resonance has been identified as the transition state resonance A,  extensively studied in the Refs. \cite{RES1}- \cite{RES5},\cite{A0}, 
to which we refer the interested reader. 
%one of the few observed experimentally \cite{EXP}, 
%labelled \e{A} in previous theoretical studies \cite{A1},\cite{A0},\cite{A2},\cite{A3},
%is a consequence of the capture of the collision partners into metastable triatomic in the van der Waals well  on the product size of the potential 
%surface.A  
Its  lifetime (i.e., the typical time the metastable 
complex exists prior to breaking up into products), $\tau=\hbar /2\text{Im}[E_\I]$ is found to be rather short, $\approx 2\times 10^{-16}$ sec. for $12 \le J \le 17$ [cf. Fig.\ref{F8}].
It is, however, premature to judge the resonance to be too short-lived to produce observable effects in the corresponding state-to state DCS. 
More important in this regard is its angular life $\phi=\hbar /2\text{Im}[J_\I]$ (i.e., the angle by which the complex rotates before breaking up into products), 
otherwise given by the product of $\tau$ with the complex's angular velocity $\om$. The latter can be considerable for a light triatomic with a large rotational 
constant $\text B\approx \R[E_n]/J(J+1)$,  and for the resonance $\text{A}$ we find $\phi$  fairly stable, varying across the chosen energy range from   $30.2^o$ to $27.9^o$ [cf. Fig.\ref{F6}].  However, even this relatively short angular life is proven to be sufficient to produce the interference 
patterns in the DCS [cf. Figs.\ref{F3} and \ref{F7}].
\newline
The resonance $\text{B}$, on the other hand, has a similar lifetime (cf. Fig.\ref{F6}), but a smaller rotational constant. 
Its rotation is slower, and the decay has little 
effect on sideway and backward scattering in Fig.\ref{F5}b and c.  Still, we found it responsible for about $20\%$ of the forward DCS in the  $62-74$ meV
energy range, as shown in Fig.\ref{F5}a.
%so that the metastable complex decays long  before completing on full rotation.
\newline
A similar behaviour is seen in the DCS of the three transitions considered here, $ j_\f=0,1,2$.
In all three cases, we found two other resonances, whose Regge trajectories were labelled  $\III$ and $\IV$. These, however, have only  minor effect on the 
DCS, since their residues are too  small, and  the patterns they produced in the DCS are practically negligible.
% (see arrows in the corresponding figures).
%\newline
%Only the resonance (I), also known as \e{B}, has a favourable combination of the life angle and the magnitude to have an affect  on the differential cross sections.
% considerably.
\newline
Therefore, the three differential cross sections considered here give a fairly clear  example of a single resonance, capable 
of  dramatically changing the nature of reactive angular scattering. 
In the lower energy range $E< 62$ meV this conclusion is, however, no longer true, as both resonances $\text{A}$ and $\text{B}$, 
are expected to play there equally important roles. We defer this case in our future work. 

In conclusion, the proposed analysis, facilitated by the \texttt{DCS\_Regge} code, can provide important insight into the reaction's mechanism. This paves the way for studying more complex chemical reactions.%%%%%%%%%%%%%%%%%
%%%%%%%%%%%%%%%%%%%%%%%%%%%%%%%%%%%%%
\section{Supplementary material}
{\it A. The unfolded amplitudes}
In the classical picture, Jacobi vector $\Rv$, drown from the centre of mass of $\text{BC}$ to $\text{A}$, rotates in the positive sense 
around the fixed direction of the total angular momentum $\bf J$. 
%The angle, swept by its projection onto the plane perpendicular to $J$ is the variable, conjugate to the $J$ \cite{CPC}.
For zero-helicity transition studied here, both the initial and final directions of $\Rv$, also lie in the plane.  
The winding angle $\varphi$, swept by the projection of $\Rv$ onto the 
plane, perpendicular to $\bf J$ is simply related to the reactive scattering angle $\tr$ as shown in Fig.\ref{F1}a.
For $0 < \varphi < \pi$ one has $\tr=\pi -\varphi$. However, the symmetry of the problem is such \cite{CPC3}, 
that a rotation by $\varphi = \pi +\tr$ also leads to the same scattering angle. 
Adding multiples of $2\pi$ one obtains all winding angles in Eq.(\ref{5a}), consistent with the chosen $\tr$.  
An angle $\va_m(\tr)$  falls into  \e{nearside} or  \e{farside} category, depending on whether $m$ is even or odd, respectively. 
\newline 
In the body-fixed frame, used in the calculation of the $S$-matrix element, $\va$ is the variable conjugate to $J$ \cite{CPL}.
For this reason, transformations from $J$- to $\va$-representation 
in Eqs.(\ref{4}) contain a simple exponential kernel $\exp(i\lm\va)$. 
The full scattering amplitude is found by \e{folding back} the \e{unfolded amplitudes}, i.e., by summing with appropriate factors
the values of $\tilde f(\va)$, or $\tilde g(\va)$, over all $\va$s consistent with the observational angle $\tr$ [cf. Eqs. (\ref{5})-(\ref{6})]. 
The procedure is illustrated schematically in Fig.\ref{F1}b.
\begin{figure}[!ht]
\includegraphics[angle=0,width=12cm, height= 5cm]{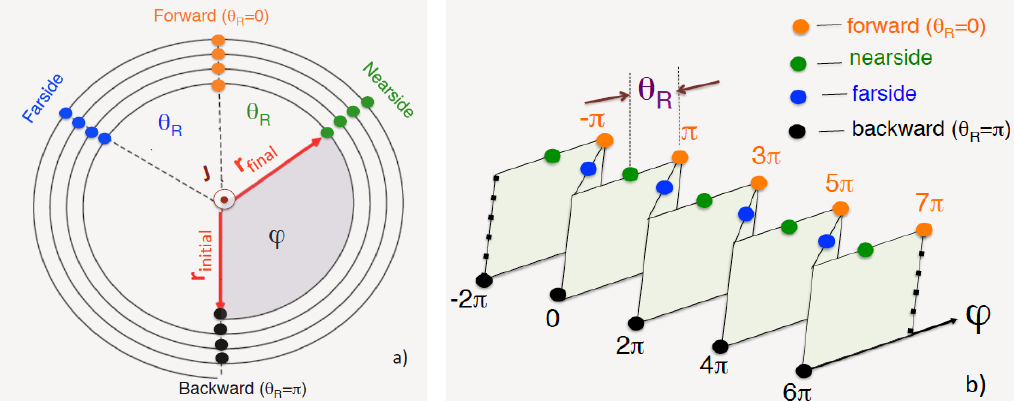}
\caption {a)
Angle $\va$ between the initial and final directions of a vector $\bf r =\Rv/|\Rv|$.
Also shown other winding angles consistent with $0<\tr<\pi$ (green and blue), $\tr=0$ (orange), and 
$\tr=\pi$ (black).
b) The values of $\varphi_m$ over which $\tilde f (\va)$ is summed in Eq.(\ref{5}) (green and blue), 
and the corresponding values for $\tilde g (\va)$ in Eqs.(\ref{6}) (orange for $\tr=0$, black for $\tr=\pi$).}
\label{F1} 
\end{figure}
%%%%%%%%%%%%%%%%%%%%%%%%%%%%%%%

{\it B. Assignment of Regge resonances.}
Figure \ref{F6A} shows the real parts of the complex energy poles
of the resonances $\text{A}$ and $\text{B}$, 
obtained by the Q-matrix analysis \cite{A1} for the Stark-Werner potential surface
(filled triangles). Also shown by the filled circles are the CE poles obtained 
in the present work by the Pad\'e reconstruction of $S^J_{3,0,0\gets 0,0,0}$ for the FXZ PES.
(This can be done with the help of the code \texttt{DCS\_Regge},  by supplying as input data $S^J_{3,0,0\gets 0,0,0}(E_j)$, 
for a fixed value of $J$, and $E_j$ $j=1,  N_j$, chosen on a suitable grid.)
The two sets of poles agree if the SW results are lowered by about  of $13$ meV, and there is a reason for that.
 The FXZ PES  \cite{OBS} is a relatively recent improvement on the SW PES, 
 developed to reproduce experimentally known exothermicity of the $F+H_2$ reaction with sufficient accuracy.
 It is also expected to accurately predict the resonance positions, which is why it was used in the present work.
 Of particular interest to us is the endothermicity of the $HF (v_f=3)$ threshold energy, 
which is reduced on the FXZ PES by  about $13$ meV (see the Table in \cite{PNAS08}).   
 It has been shown \cite{A0}, \cite{A1}, \cite{RES3}  earlier that the resonance energies are well predicted 
 by the energies  of the quasi bound states on the vibrationally adiabatic $FH+H$ potential curves. For this reason, we adjust the SW results shown in Fig.\ref{F6A} by $13$ meV to create Fig.\ref{F6}b. 
The resulting good agreement, evident in Fig.\ref{F6}b, demonstrates that the CE poles, obtained for the Regge trajectory 
I (blue circles), correspond to the exit channel resonance B. By the same token, the CE poles from the Regge 
trajectory II, can be attributed to the transition state resonance A. 
%%%%%%%%%%%%%%%%%%%%%%%%%%%%%%%
\begin{figure}[!ht]
\includegraphics[angle=0,width=12cm, height= 7cm]{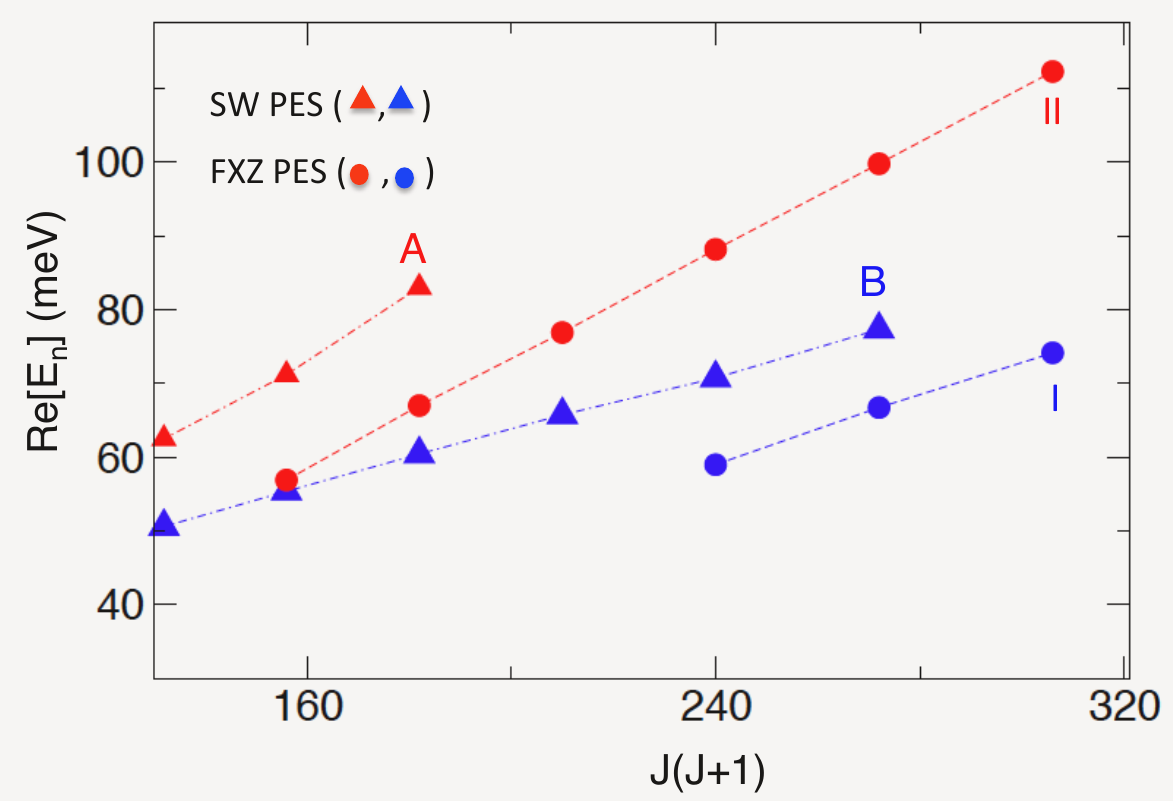}
\caption { Comparison of
the CE poles, obtained for the Stark-Werner (SW) PES
by the Q-matrix method of \cite{A1} (triangles) with 
 the poles, obtained for the FXZ PES by the Pad\'e reconstruction 
 (circles, present work). }
 %The dashed lines show linear fits to the SW data.}
%b) Same as (a), but with the SW poles shifted down by 12 meV.}
\label{F6A} 
\end{figure}
%%%%%%%%%%%%%%%%%%%%%%%%%%%%%%%%%

{\it C. Transitions $(0,0,0)\to (3,1,0)$  and  $(0,0,0)\to (3,2,0)$}.
The amplitudes $\tilde{f}(\va,E)$ and $\tilde{g}(\va,E)$, shown in Fig.\ref{F9}, 
exhibit in the region $\va \ge \pi$ decaying tails, similar to those seen in Figs.\ref{1}b and c.
As in the case of the $(0,0,0)\to(3,0,0)$ we attribute them to the Regge trajectory $\II$ in Fig.\ref{F8},
corresponding to the resonance $\text{A}$ of Sect.II $B$. 
\begin{figure}[!ht]
\includegraphics[angle=0,width=12cm, height= 10cm]{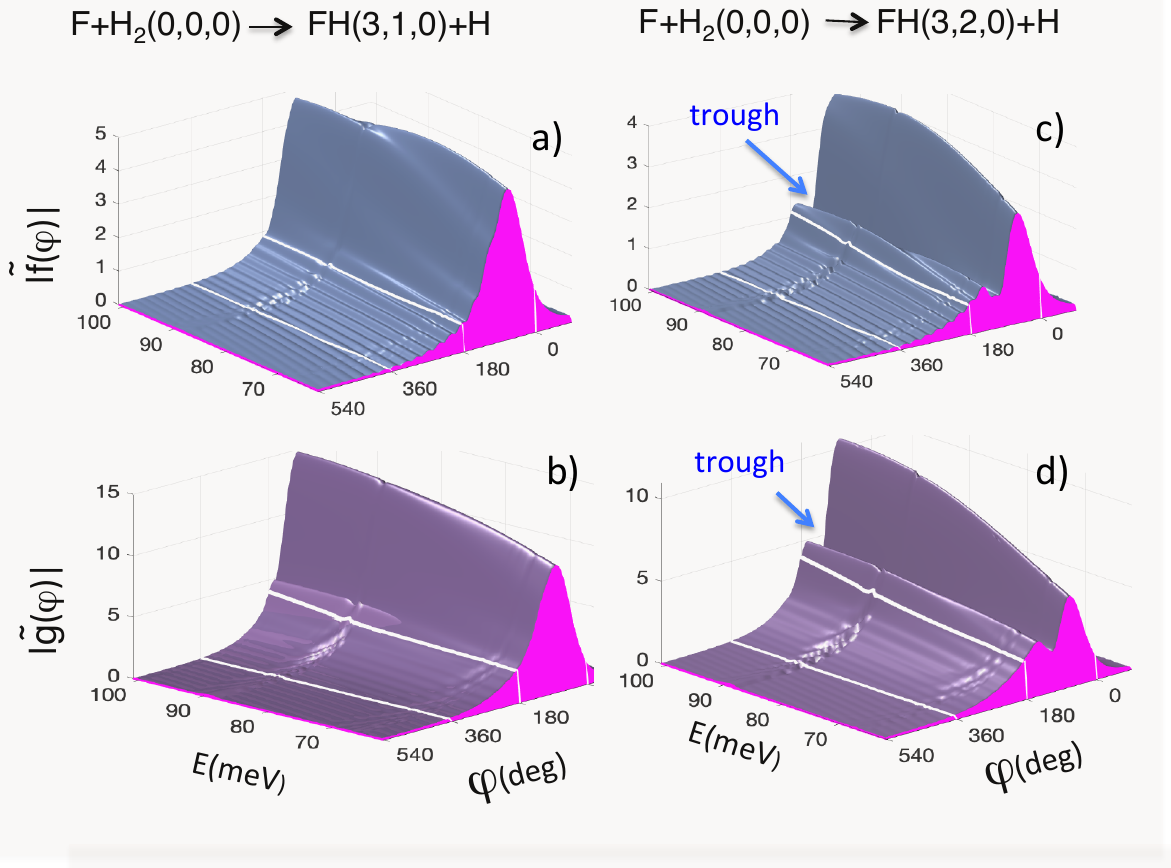}
\caption {a) The modulus of the unfolded amplitude $\tilde{f}(\va,E)$ for the $(0,0,0)\to (3,1,0)$ transition.
\newline
b) The modulus of the unfolded amplitude $\tilde{g}(\va,E)$ for the $(0,0,0)\to (3,1,0)$ transition.
\newline
c) Same as (a) but for the $(0,0,0)\to (3,2,0)$ transition.
c) Same as (b) but for the $(0,0,0)\to (3,2,0)$ transition.}
\label{F9} 
\end{figure}
%%%%%%%%%%%%%%%%%%%%%%%%%%%%%
The zeroes of $S_{3,2,0\gets 0,0,0}(J,E)$ in the complex $J$-plane are shown in Fig.\ref{F10}. 
One notes a regular zero trajectory, responsible for the trough in Figs.\ref{9}c and d (highlighted).
Also clearly visible are the zeroes which accompany Regge trajectories $\III$ and $\IV$ shown in Fig.\ref{F8} (note that no zeroes accompany the pole trajectories I and II).
\begin{figure}[!ht]
\includegraphics[angle=0,width=12cm, height= 8cm]{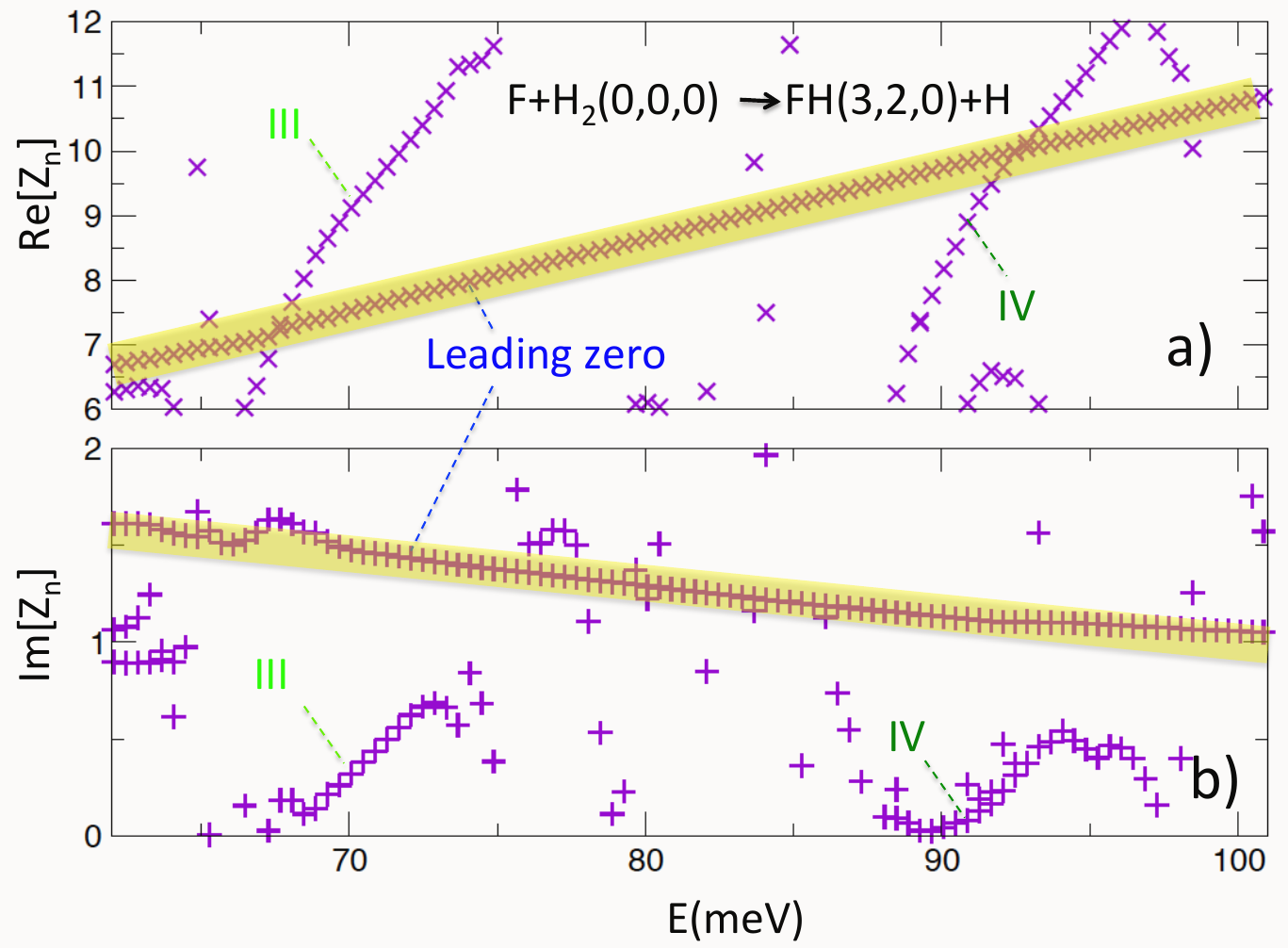}
\caption {a)
Real parts of the CAM zeroes (crosses) of  $S_{3,2,0\gets 0,0,0}$ vs. $E$. 
b) Same as (a) but for the zeroes' imaginary parts.}
%Highlighted is the trajectory, responsible fro the maximum in Fig.\ref{F7}e, and a minimum 
i%n  Fig.\ref{F11}f.} 
\label{F10} 
\end{figure}
%%%%%%%%%%%%%%%%%%%%%%%%%%%

 \section*{Acknowledgements}
D.S. acknowledges financial support by the Grant PID2021-126273NB-I00 funded by MICINN/AEI/10.13039/501100011033 and by "{ERDF A way of making Europe}", as well as by the Basque Government Grant No. IT1470-22.
D.DF acknowledges the CINECA award under the ISCRA initiative for the availability of high performance computing resources and support.
E.A. acknowledges the financial support by MICIU/AEI/10.13039/501100011033 and FEDER, UE through BCAM Severo Ochoa accreditation CEX2021-001142-S / MICIU/ AEI / 10.13039/501100011033; PLAN COMPLEMENTARIO MATERIALES AVANZADOS 2022-2025, PROYECTO No:1101288, and grant PID2022-136585NB-C22; as well as by the Basque Government through ELKARTEK program under Grants KK-2023/00017, KK-2024/00062 and the BERC 2022-2025 program.%E.A. acknowledges the financial support by the Ministerio de Ciencia y Innovaci\'on (MICINN, AEI) of the Spanish Government through BCAM Severo Ochoa accreditation CEX2021-001142-S and PID2019-104927GB-C22, PID2022-136585NB-C22 grants, as well as by the Basque Government through the BERC 2022-2025 Program, IKUR Program, ELKARTEK Programme (grants KK-2022/00006, KK-2023/00016).

\end{document}